\documentclass[superscriptaddress,aps,prl,floats,twocolumn,twoside,floatfix]{revtex4}
\usepackage{epsfig}
\usepackage{amsmath}
\usepackage{amssymb}

\newcommand{\be}{\begin{equation}}
\newcommand{\ee}{\end{equation}}
\newcommand{\bea}{\begin{eqnarray}}
\newcommand{\eea}{\end{eqnarray}}
\begin{document}
\title{Answer to Comment on ``Ultrametricity in the Edwards-Anderson Model'' http://arxiv.org/abs/0709.0894}  
\author{Pierluigi Contucci}
\affiliation{
Universit\`{a} di Bologna, Piazza di Porta S.Donato 5, 40127 Bologna, Italy}
\author{Cristian Giardin\`a}
\affiliation{Eindhoven University and EURANDOM, P.O. Box 513 - 5600 MB Eindhoven, The Netherlands}
\author{Claudio Giberti}
\affiliation{ Universit\`a di Modena e Reggio Emilia, via G.
Amendola 2 -Pad. Morselli- 42100 Reggio Emilia, Italy}
\author{Giorgio Parisi}
\affiliation{
Universit\`a La Sapienza, Piazzale A. Moro, Roma, Italy}
\author{Cecilia Vernia}
\affiliation{
Universit\`{a} di Modena e Reggio Emilia, via Campi 213/B, 41100 Modena, Italy}
\maketitle
In a recent comment \cite{JK} to our Letter \cite{CGGPV} T.Jorg and F.Krzakala
have investigated the properties of 2-dimensional Edwards-Anderson (EA) model
and found, by numerical methods, the interesting result that some ultrametric features
hold for the link overlap probability distributions for square lattices
of side $16$ and $32$. Namely,
\bea
\label{uno}
\rho_X(x) & = & \delta(x)
\nonumber \\
\rho_Y(y) & = & \frac{1}{4}\delta(y) + \frac{3}{2}\theta(y)\int_{y}^1 P(a)P(a-y)da
\eea
where $X$ (resp. $Y$) is the difference between the
medium and the smallest (resp. the difference between the largest and the medium)
for a triplet of link overlaps sampled with respect to the equilibrium quenched measure.
Their results are illustrated in figure 1 of \cite{JK}. Since at positive temperature in two
dimensions the RSB picture cannot hold they conclude that the results presented in \cite{CGGPV}
are not sufficient to dismiss the droplet picture in the 3-dimensional EA model.

Our answer can be summarized as follows.

\begin{figure}
\setlength{\unitlength}{1cm}
          \centering
                \includegraphics[width=9cm,height=5cm]{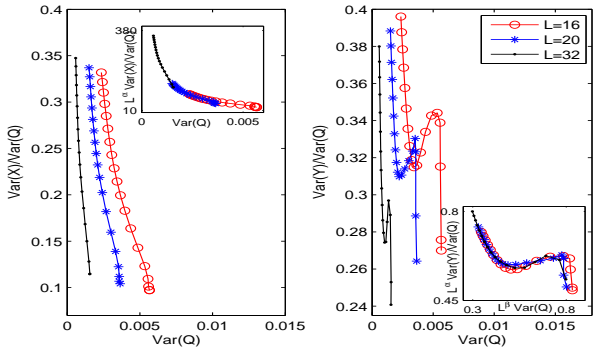}
               \caption{Normalized variances of the two random variables $X$ (left) and
               $Y$ (right) as a function of $Var(Q)$ for 2D $+/-1$. The left inset shows the scaling law
               $L^{\alpha}Var(X)/Var(Q)$ for $\alpha=2$ and the right inset the scaling law
               $L^{\alpha}Var(Y)/Var(Q)$ as a function of $L^{\beta}Var(Q)$ for $\alpha=0.22$
               and $\beta=1.8$. In both cases the scaled normalized variances are $L$-independent.}
\label{fig1}
\end{figure}

\noindent
The conclusions obtained in \cite{CGGPV} are {\it mainly} based on the analysis of the scaling
properties of the variance for the two random variables $X$ and $Y$. The statement in favor of
ultrametricity that we made is in fact based on the observation (see Fig. 1 and 3 in \cite{CGGPV})
that for increasing volumes the variance of the variable $X$ is shrinking to zero with a
suitable scaling law while the variance of the variable $Y$ is not. The subsequent analysis
of the distribution shape for some finite volume (see Fig. 2 in \cite{CGGPV}) was indeed proposed as a further
support of the main result. Since the study in \cite{JK} is only concerned with a finite
volume analysis of the overlap distributions with NO asymptotic analysis it cannot be used
to weaken the conclusions obtained in \cite{CGGPV}. In order to parallel the approach followed
in \cite{CGGPV} one should have performed in fact, prior to analyse the probability
distribution for some finite volume, the asymptotic behaviour of the variances of the variables $X$ and $Y$.
This can be done indeed with a modest computational effort and gives the result shown in Fig. \ref{fig1}. One
immediately sees that both the variances are shrinking to zero and by consequence ultrametricity doesn't hold.
This shows that the method developed in \cite{CGGPV} is robust.

Still the observation that in $d=2$ the overlap distribution shape stays, for the volumes considered,
close to that predicted by the RSB picture is interesting and deserves a proper explanation.

Here we notice that in the $(T,d)$ plane - dimension vs. temperature - there is a curve
which separates in the thermodynamic limit the region with broken symmetry (the upper one) from the
paramagnetic one. At $T=0$ the curve crosses the $d$ axis on the lower critical dimension
$d_l= 2.5$ \cite{B} and it grows for positive temperatures $T>0$. According to the RSB picture \cite{MPV}
the upper region is characterized by a spin glass phase with an ultrametric overlap distribution.
However, for a finite volume system, if one is outside the spin glass region but close enough to
the critical curve in the $(T,d)$ plane one might still observe some features of an ultrametric
overlap distribution.

The point investigated in \cite{JK} $(T=0.2,d=2)$ is just below the critical curve and
not far enough to observe, for the volumes they investigate which in $d=2$ are quite
small, the thermodynamic properties.

To support our claim it is enough to investigate the point $d=1$ and $T=0$. Since we
are now really away from the critical curve ultrametricity cannot hold anymore.
At zero temperature the relevant states in the Gibbs measure are only ground states,
which for a frustrated closed chain are kink and anti-kink (we disregard non frustrated
disorder samples which are obviously ferromagnetic).

We analyzed the probability distribution for a triplet of {\em standard} overlap,
since in $d=1$ the link overlap is 1 with probability 1.
As one expects the first signal of violation of ultrametricity can be detected studying the behaviour of
the random variable $S= sign(q_{1,2}\,q_{2,3}\,q_{3,1})$ (see also \cite{NS}).
An explicit computation shows that the quenched expectation $<S>=1/2$, while for an ultrametric
topology, as the one predicted by RSB theory, one should have $<S>=1$. Moreover, for the distribution
of $\tilde X = \tilde q_{med} - \tilde q_{min}$ and $\tilde Y = \tilde q_{max} - \tilde q_{med}$,
where
\bea
\tilde q_{max} & = & max (|q_{1,2}|,|q_{2,3}|,|q_{3,1}|)\nonumber \\
\tilde q_{med} & = & med (|q_{1,2}|,|q_{2,3}|,|q_{3,1}|)\nonumber \\
\tilde q_{min} & = & S \,min (|q_{1,2}|,|q_{2,3}|,|q_{3,1}|)
\eea
a simple numerical simulation sees that the plots are
aligned along the lines
\bea
\rho_{\tilde X}(x) & = & \left\{\begin{array}{ll}
-\frac92 x + 3 & \textrm{for $0\le x\le \frac23$}\\
0 & \textrm{for $\frac23 < x \le 1$}
\end{array}\right.
\nonumber \\
\rho_{\tilde Y}(y) & = & -2y + 2
\eea
as also a direct analytical computation shows in the large volume limit.
Clearly the previous formulas cannot satisfy the RSB ultrametric relation of Eq. (\ref{uno}) .

In conclusion the argument of \cite{JK} cannot be used to weaken the result presented in \cite{CGGPV}
which was based mainly on identifying the scale law that governs the approach to
ultrametricity. The same method used in \cite{CGGPV} and properly applied to the two
dimensional case reveals in fact the expected lack of RSB picture at positive temperature.
The numerical result presented in \cite{JK} is interesting only because
it shows that for small volumes some ultrametric features may persist if the
system is investigated outside but close enough to the spin glass region. The
{\it apparent} ultrametricity observed in \cite{JK} is due to the closeness
of their simulation to the critical line in the plane $(T,d)$. Moving away
from the line, for instance performing the analysis in $(0,1)$
the disappearance of ultrametricity is immediately seen also in finite volume systems.

{\bf Acknowledgments.} We want to thank T.Jorg and F.Krzakala for a useful correspondence.


\begin{thebibliography}{99}
%
\bibitem{JK} T.Jorg and F.Krzakala
{\em Comment on "Ultrametricity in the Edwards-Anderson Model"}
{\em http://arxiv.org/abs/0709.0894}
%
\bibitem{CGGPV} P.Contucci, C. Giardin\`a, C. Giberti, G.Parisi, C.Vernia,
{\em Phys. Rev. Lett.} {\bf 99}, 057206 (2007)
%
\bibitem{B} S. Boettcher
{Phys. Rev. Lett.} {\bf 95}, 197205 (2005).
%
\bibitem{MPV}
M. Mezard, G. Parisi, M.A. Virasoro,
{\em Spin Glass Theory and Beyond}
World Scientific, Singapore (1987).
%
\bibitem{NS}
C. Newman, D. Stein,
{\em Percolation in the Sherrington-Kirkpatrick spin glass}
{\em http://arxiv.org/0710.1399}


\end{thebibliography}
\end{document}